\documentclass[preprint,secnumarabic,amssymb,bibnotes, footinbib, aip, jap]{revtex4-1}

\usepackage{graphicx}
\usepackage{amsfonts,amsmath,amssymb}

\newcommand{\atan}{\arctan}
\begin{document}
\title{Current induced domain wall dynamics in the presence of spin orbit torques}
\author{O. Boulle, L. D. Buda-Prejbeanu, E. Ju\'e,  I.M. Miron and G. Gaudin} 
\affiliation{SPINTEC, CEA/CNRS/UJF/INPG, INAC, 38054 Grenoble Cedex 9, France}

\date{\today}%
\begin{abstract}
Current induced domain wall (DW) motion in perpendicularly magnetized nanostripes in the presence of spin orbit torques is studied. We show using micromagnetic simulations that the direction  of the current induced DW motion and the associated DW velocity depend on the relative  values of the field like torque (FLT) and the Slonczewski like torques (SLT). The results are well explained by a collective coordinate model which  is used to draw a phase diagram of the DW dynamics as a function of the FLT and the SLT. We   show that  a large increase in the DW velocity can be reached by a proper tuning of both torques.
\end{abstract}

\maketitle
The dynamics of magnetic domain walls  (DW) induced by a spin polarized current has attracted a large effort of research during the last ten years motivated not only by promising devices in the field of magnetic storage and logics~\cite{Parkin08S} but also by the wealth of the physics involved. The principle behind current induced DW motion is the spin transfer effect  where the spin current going through the DW is transferred to the DW magnetization leading to spin transfer torque and DW motion in the carrier direction~\cite{Berger78JAP}. Whereas first experiments in soft in-plane magnetized stripes were relatively well described by this scheme, it was shown later that the spin transfer effect was much more efficient in ultrathin out-of-plane magnetized multilayers, such as Pt/Co/Pt~\cite{Boulle11MSaERR} or Pt/Co/AlOx multilayers~\cite{Miron11NMa}  meaning that other effects were at stack. This was first interpreted as the result of very large additional non-adiabatic effects due to the incomplete absorption of the spin current~\cite{Thiaville05EL}, but the physical mechanisms behind were elusive~\cite{Boulle11MSaERR}. In addition, puzzling experiments were reported, for example in asymmetric  Pt(4~nm)/Co(0.6~nm)/Pt(2~nm) multilayers, where the    DW is moved    in the direction or opposite to the current direction  when reversing the position of both Pt layers in the stack, in contradiction with the standard spin transfer mechanism~\cite{Lavrijsen11APL,Haazen13NM}.  Recently, it was shown that the large spin orbit coupling due the presence of heavy metal (such as Pt or Ta) as well as the inversion asymmetry due to the interfaces can lead to additional current induced spin-orbit torques~\cite{Miron11N,Liu12S,Garello13NN}. Two types of torques have been identified: (1) a field like  torque (FLT) $\mathbf{m}\times J H_{FL} \mathbf{u_y}$  similar to the one exerted by an in-plane magnetic field $J H_{FL}\mathbf{u_y}$ oriented perpendicularly to the current direction. (see Fig.~\ref{Fig2}) (2) a Slonczewski like torque (SLT)   $-\gamma_0 JH_{SL}\mathbf{m}\times(\mathbf{m}\times\mathbf{u_y})$. 
 Recent experiments have underlined that these torques may actually explain the apparent  very large efficiency of the spin transfer effect reported in these systems~\cite{Haazen13NM,Ryu13NN,Emori13NM}.  

In this paper, we study how the FLT and the SLT  affect the current induced DW dynamics.  Using micromagnetic simulations,   we show that depending on their relative values, the DW can move in one direction or the other with respect to the current direction.  The results of micromagnetic simulations are well reproduced by a   collective coordinate model (CCM) taking into account both torques.   This model is used to predict a phase diagram of the DW direction and velocity as a function of   the FLT and SLT    and allows the identification of the torque  conditions for maximum velocity. 

We consider a perpendicularly magnetized stripe with a width of 100 nm.  Micromagnetic simulations are based on  the Landau-Lifschitz-Gilbert equation to which the current induced torques have been added:
\begin{equation}\label{LLG}
\frac{\partial \textbf{m}}{\partial t}= -\frac{\gamma_0}{\mu_0M_s} \frac{\delta E}{\delta\textbf{m}}\times \textbf{m} + \alpha \textbf{m} \times \frac{\partial \textbf{m}}{\partial t} -u\frac{\partial \mathbf{m}}{\partial x}+\beta u \mathbf{m}\times\frac{\partial \mathbf{m}}{\partial x}-\gamma_0 \mathbf{m}\times\mathbf{m}\times H_{SL}J\mathbf{u_y}+\gamma_0\mathbf{m}\times JH_{FL} \mathbf{u_y}
\end{equation}
where $\gamma_0=\mu_0\gamma$ with $\gamma$ the gyromagnetic ratio, $E$ the energy density and $M_s$ the saturation magnetization. The third term is the adiabatic spin-transfer torque where $u=JPg\mu_B/(2eM_s)$,  $\mu_{B}$   the Bohr magneton,  $J$ the current density, $M_s$ the saturation magnetization, $P$  the current spin polarisation. The fourth term is the non-adiabatic torque described by the dimensionless parameter  $\beta$~\cite{Thiaville05EL}. 
For the micromagnetic simulations, the following parameters have been used: the exchange constant $A=10^{-11}$~A/m, the anisotropy constant $K^0_{an}=1.25\times10^6$~J/m$^3$,  $M_s=1.1\times10^6$~A/m, the damping parameter $\alpha=0.5$,  $\beta=0$, $P=1$. The thickness of the magnetic layer is 0.6~nm. 3D micromagnetic simulations were carried out using a homemade code~\cite{Buda02CMS}. 
Note that we do  not consider  the Dzyaloshinskii-Moryia interaction~\cite{Thiaville12EL,Boulle13,Ryu13NN,Emori13NM} so that  a Bloch DW equilibrium configuration is observed.  
\begin{figure}[!h]
	\centering
		\includegraphics[width=1\textwidth]{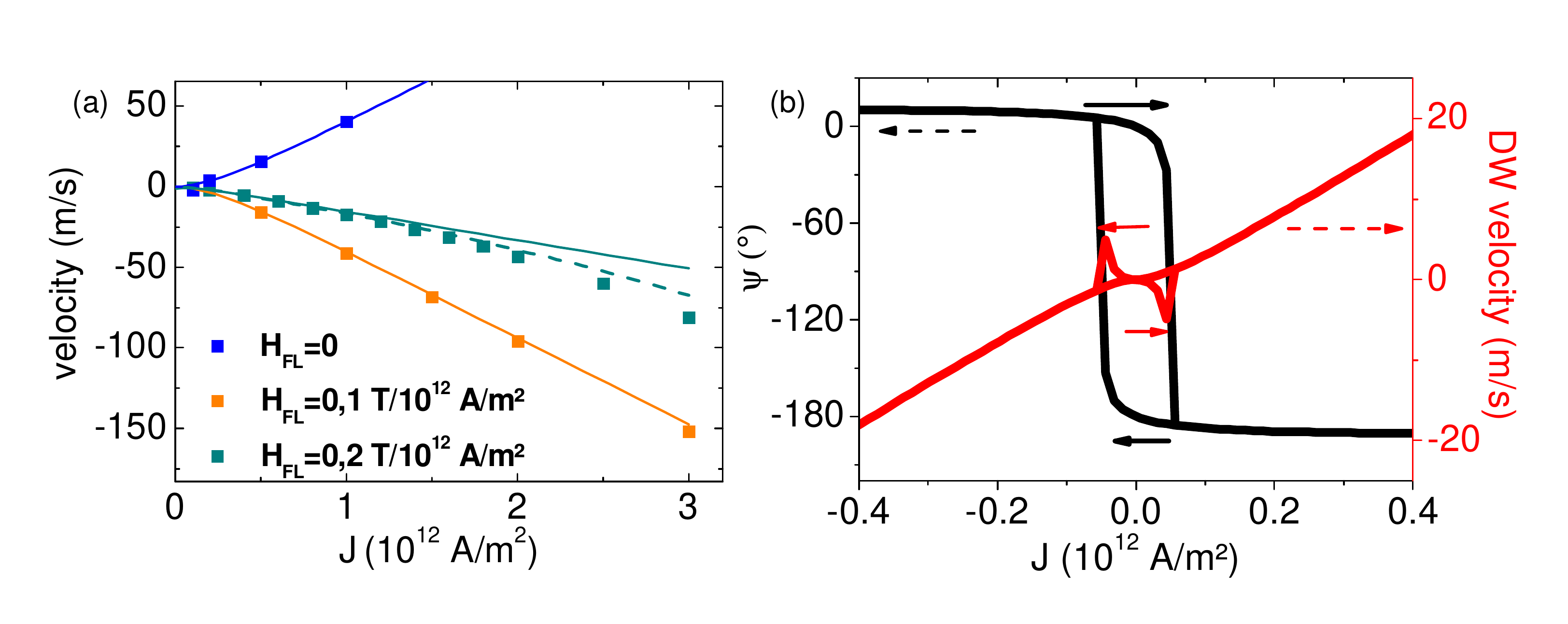}
	\caption{ (a) DW velocity as a function of the current density for  $H_{SL}=0.025$ T/($10^{12}$~A/m$^2$)  and  different values of the FLT deduced from micromagnetic simulations (dots), the standard CCM (continous line) and the extended CCM taking into account the DW deformation induced by the FLT (dashed line). The DW demagnetizing field $H_k=33$~mT is used for the CCM simulations.(b) DW velocity and DW angle $\psi$ as a function of $J$ ($H_{FL}=0$ and $H_{SL}=0.1$~T/($10^{12}$~A/m$^2$)) calculated from the CCM. }
	\label{Fig1}
\end{figure}
Fig.~\ref{Fig1}(a) shows the results of micromagnetic simulations (dots)  of the DW velocity as a function of the current density for  $H_{SL}=0.025$~T/($10^{12}$~A/m$^2$),  and different values of the FLT. The FLT strongly affects the current induced DW motion: depending on the value of the FLT, the DW velocity is positive or negative, meaning the DW moves in the direction or opposite to the current direction, and the DW velocity amplitude depends non-monotically on the FLT.

\begin{figure}[!h]
	\centering
		\includegraphics[width=0.7\textwidth]{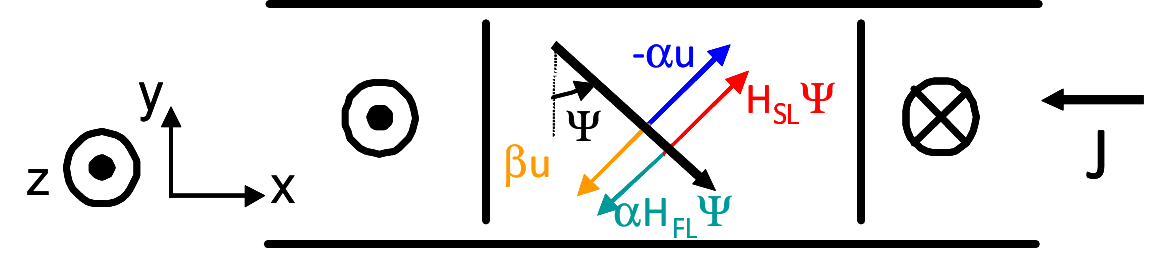}
	\caption{  Schematic representation of the different current induced torques acting on the DW magnetization.}
	\label{Fig2}
\end{figure}

To better understand these results, we consider a collective coordinate model (CCM) which assumes that the DW keeps  its static structure during its motion whereas the forces inducing the dynamics are first order correction affecting only the DW position $q$ and the DW angle $\psi$  (see Fig.~\ref{Fig2})~\cite{Slonczewski79,Thiaville05EL}.  

We first assume small current density so that  the DW structure and the magnetization in the domain are  little affected by the spin orbit torques and one can assume a standard Bloch profile. The polar and azimuthal angles  $\theta$ and $\varphi$ are then assumed as  $\theta=2\atan(\exp\{[x-q]/\Delta(\psi)\})$ and   $\varphi=\psi-\pi/2$ with $\psi$ constant.    Here $\Delta(\psi)= \sqrt{A/\kappa}$  is the DW width with $\kappa=(K_{an}+K\sin^2 \psi)^{1/2}$ where $K_{an}=K_{an}^0-\mu_0M_s^2$,   $K=\mu_0M_s H_k/2$ with $H_k$ the DW demagnetizing field.  The integration of Eq.~\ref{LLG} over this DW profile leads to  the following   equations:
\begin{eqnarray}\label{1Dmodel1}
\dot{\psi}+\frac{\alpha\dot{q}}{\Delta}=\frac{\beta u }{\Delta}+\gamma_0 J H_{SL}\frac{\pi}{2}\sin\psi\\
\frac{\dot{q}}{\Delta}-\alpha\dot{\psi}=\frac{\gamma_0 H_k}{2}\sin2\psi+\frac{u}{\Delta}+\gamma_0 H_{FL}J\frac{\pi}{2}\sin\psi
\label{1Dmodel2}
\end{eqnarray}
The SLT   enters the equations as an additional force on the DW  which is  proportional to  $\sin\psi$, i.e to the component of the DW magnetization along the current direction. It   is thus zero for a perfect Bloch configuration whereas it is maximal for a N\'eel configuration. The effect of the SLT on the DW can be seen as an effective non-adiabatic parameter $\beta_{SL}\sin\psi$ with $\beta_{SL}=\gamma_0 H_{SL}\Delta\pi eM_s/(g\mu_BP)$. For the value of $H_{SL}\sim0.07$~T/($10^{12}$~A/m$^2$) measured in Pt~\cite{Garello13NN}, one obtains a  large  value   $\beta_{SL}\sim2$.  The SLT can thus account for the large $\beta$ value reported in the literature in perpendicularly magnetized nanostripes~\cite{Boulle11MSaERR}. A simple expression of the DW velocity can be deduced from Eq.~(\ref{1Dmodel1}-\ref{1Dmodel2}) in the steady state regime ($\dot{\psi}=0$):
\begin{equation}
\label{V1D}
	v=\frac{\beta u }{\alpha}+\frac{\gamma_0 H_{SL}J\Delta}{\alpha}\frac{\pi}{2}\sin\psi
\end{equation}


The  DW angle $\psi$ will thus determine the direction and the amplitude of the DW velocity. The effect of the SLT is in fact similar to  an easy axis magnetic field  $\mathbf{H_{SL}}=\mathbf{m}\times JH_{SL}\mathbf{u_y}=\sin\psi JH_{SL}\mathbf{u_z}$.   For the configuration of Fig.~\ref{Fig2} and $J>0$, a positive (resp. negative) $\sin\psi$ leads to $\mathbf{H_{SL}}$   aligned upward (resp. downward) and thus moves the DW in the electron (resp. current)  direction.  

The steady state value of $\psi$ is   the results of a balance between the different in-plane component of the current induced torques (see Fig.~\ref{Fig2}). 
This balance can be simply expressed in the steady state regime ($\dot{\psi}=0$)  by considering small angles $\psi$ close to 0 or $\pi$:
\begin{eqnarray} 
\psi=\frac{u(\beta-\alpha)/\Delta}{\alpha\gamma_0 H_k+\epsilon\frac{\pi}{2}\gamma_0 J(\alpha H_{FL}-H_{SL})} (+\pi)\\
v=\frac{\beta u }{\alpha}+\frac{\pi}{2}\frac{\gamma_0 H_{SL}J}{\alpha}\frac{u(\beta-\alpha)}{\frac{\pi}{2}\gamma_0 J(\alpha H_{FL}-H_{SL})+\epsilon\alpha\gamma_0 H_k}
\end{eqnarray}
 where $\epsilon=1$ (resp. $\epsilon=-1$) for $\psi\sim0$ (resp. $\psi\sim\pi$).
The direction of the DW motion is thus set by the relative values of $\beta$ and $\alpha$ on the one hand, and by the relative value $H_{SL}$ and $\alpha H_{FL}$ on the other hand. One can show there are actually two steady state stable positions for the angle $\psi$ close to 0 and $\pi$ and at low current density, $\psi$ can be switched hysteretically between these two positions when sweeping current (see Fig.~\ref{Fig1}(b)). 
We show on Fig.~\ref{Fig1}(a) (continuous line), the DW velocity predicted by this CCM. A good agreement is obtained for $H_{FL}=0$ or $H_{FL}=0.1$ T/($10^{12}$~A/m$^2$) but the agreement is less satisfactory at larger current density for $H_{FL}=0.2$ T/($10^{12}$~A/m$^2$). In this case, the large transverse magnetic field induced by the FLT affects the domain   and DW structure and the simple assumption of a Bloch DW structure does not hold. 

To account for the large transverse field induced by the current injection $H_t=JH_{FL}$ , we consider a    CCM  which assumes   a more complicated  DW structure where  the domain and DW deformation induced by $H_t$ is taken into account~\cite{Hubert74}. 
To describe the current induced DW dynamics for such a DW profile, a Lagrangian approach is considered (see Ref.~\cite{Hubert74,Boulle12JAP,Boulle13}). 
The Lagrange-Rayleigh equation then leads to the following equations :
\begin{eqnarray}
\label{Eq:vsteady1}
\alpha\frac{\sigma}{4A}(\dot{q}-\frac{\beta}{\alpha}u)-\gamma_0 H_{SL}J\varphi_0 \sin\theta_0\cos(\theta_0-\psi)+\frac{\partial f_k}{\partial\psi}\dot{\psi}  =   0\\
\label{Eq:Vsteady2}
\frac{\partial f_k}{\partial\psi}(\dot{q}-u)-\frac{\alpha \sigma}{4A}\left[\frac{1}{k_0^2}+\frac{1}{3}\left(\frac{\partial 1/k_0}{\partial \psi}\right)^2\left(\pi^2-\varphi_0^2-\frac{2\varphi_0^2}{1-\varphi_0\mathrm{cotan}\varphi_0}\right)\right]\dot{\psi} =  \frac{\gamma_0}{2M_s}\frac{\partial \sigma}{\partial \psi}
\end{eqnarray}
Here $\sigma$ is the DW energy density, $\theta_0=\arccos(-h\sin\psi)$, $\varphi_0=\arctan[\frac{\sqrt{1-h^2}}{h\cos\psi}]$ with $h=J H_{FL}/H_{an}$; $k_0=\sin\varphi_0\sqrt{(K_{an}+K\sin^2\psi)/A}$, $f_k(\psi,h)$ is a dimensionless parameter   defined in Ref.~\cite{Hubert74,Boulle12JAP}. From Eq.~\ref{Eq:vsteady1}, one can easily derive the   DW velocity in the steady state regime ($\dot{\psi}=0$):
\begin{equation}
\label{Vsteady2}
	\dot q=\beta u/\alpha + \frac{4A}{\sigma\alpha}\gamma_0 H_{SL}J\varphi_0\sin\theta_0\cos(\theta_0-\psi)
\end{equation}
If one compares Eq.~\ref{Vsteady2} to Eq.~\ref{V1D}, one can see that $H_t$ only affects the DW velocity induced by the  SLT and this in two ways. First, the DW width in Eq.~\ref{V1D} is replaced by an ``effective" DW width $4A/\sigma$, which is increased   by $H_t$. Second, $H_t$ changes the steady state angle $\psi$ and thus the DW velocity. These two elements   lead  to an increase of the DW velocity as compared to the standard CCM model.
We plot on Fig.~\ref{Fig1}(a) (dotted lines), the prediction of  the model in the case of a high FLT $H_{FL}=0.2~T/(10^{12}$~A/m$^2$). One can see that the model allows a better agreement with the micromagnetic simulations compared to the standard CCM model, in particular for high current density where $H_t$  is large. However, one limitation  of our model is   that it does not take into account the deformation of the domain induced by the SLT, which may be relevant at high current density and large values of the SLT such that $J H_{SL}\sim H_{an}$.

\begin{figure}[!h]
	\centering
		\includegraphics[width=0.5\textwidth]{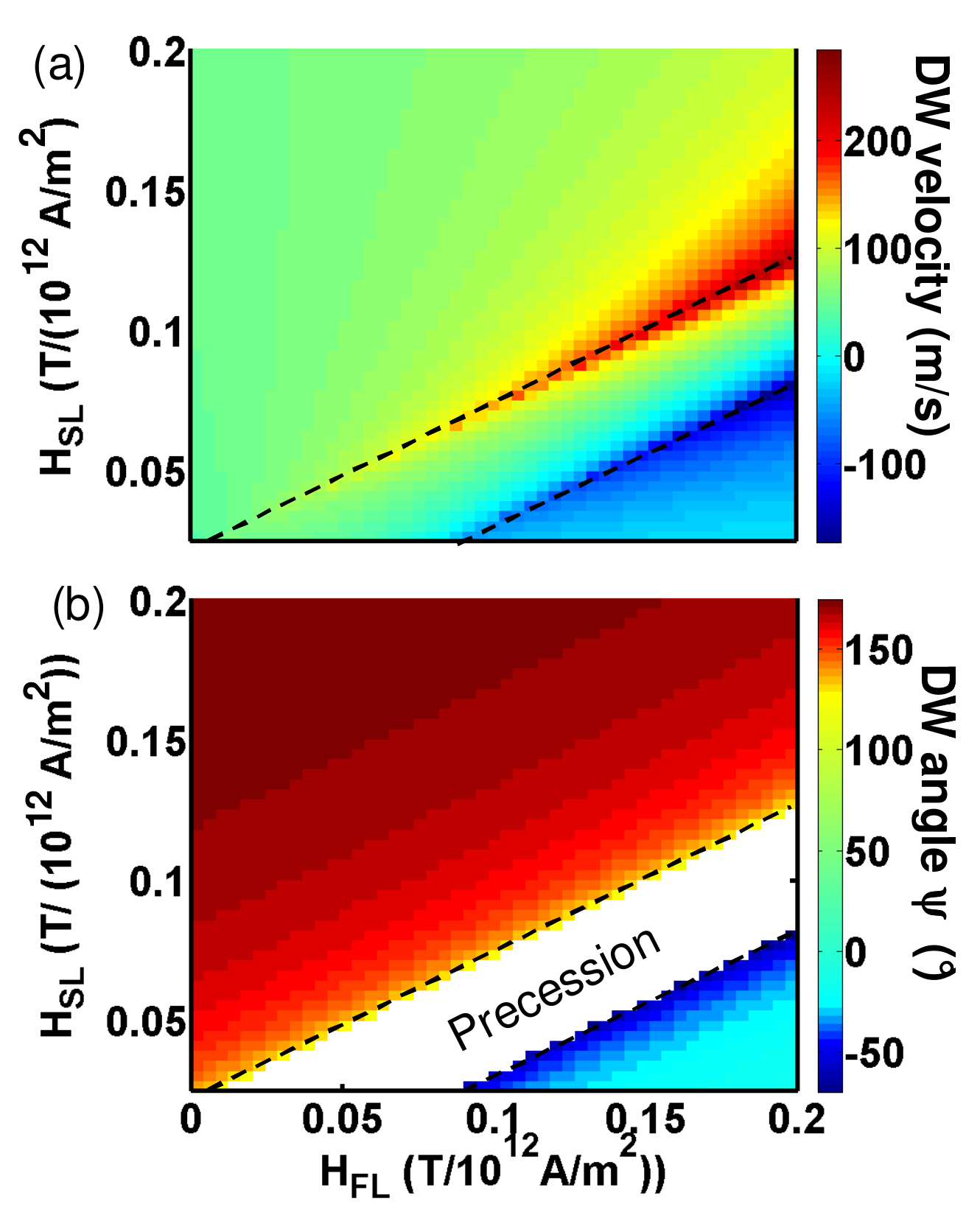}
	\caption{(a) DW velocity  and  (b) steady state DW angle $\psi$  in color scale as a function of the SLT and the FLT, obtained by solving numerically Eq.~(\ref{Eq:vsteady1}-\ref{Eq:Vsteady2}) for $J=10^{12}$~A/m$^2$. The black lines mark out the region associated with precessional motion. }
	\label{Fig3}
\end{figure}

This extended model can be used to draw a phase diagram of the DW   dynamics  as a function of   the FLT and SLT. Fig.~\ref{Fig3} shows (a) the DW velocity and (b) the DW angle $\psi$ in color scale as a function of the SLT and the FLT for $J=10^{12}$~A/m$^2$, obtained by solving numerically Eq.~(\ref{Eq:vsteady1}-\ref{Eq:Vsteady2}). One can note that the direction of the DW motion depends on the relative values of the SLT and FLT:   on the top left of the diagram, for large  SLT or  low FLT, the DW velocity is positive  whereas on the lower right corner, for large FLT and low SLT, the DW velocity is negative. This change in the direction of the DW motion goes with a switching of the DW chirality with $\pi/2<\psi<\pi$ ($\sin\psi>0$) for the positive velocity region and $-\pi/2<\psi<0$ ($\sin\psi<0$) for the negative velocity region. As expected from Eq.~\ref{V1D}, the direction of the DW motion is thus determined by the sign of $\sin\psi$. In these two regions, the DW dynamics stays  in the steady state regime whereas in between, a small region  with   precessional regime is observed. Interestingly, the largest DW velocity are predicted at the border of the precessional regime, and correspond to $\psi$ close to $\pm\pi/2$ where the SLT is maximum. For this particular condition, the SLT and FLT nearly compensate so that large angle $\psi$ and thus large SLT can be reached. One can show that these borders with maximal velocity corresponds actually to the condition $H_{SL}\approx\alpha H_{FL}\pm(\alpha P\hbar)(\Delta e M_s)$. A large DW velocity can thus be reached by a proper tuning of the SLT and FLT. Experimentally, whereas the SLT and the FLT due to the spin orbit coupling are more related to the intrinsic properties of the material, the Oersted field generated by the current flowing in the metallic layers surrounding the magnetic layer  also leads to a FLT.  For a given current density, its amplitude can be easily modified by playing on the width and thickness of the metallic layers. Finally,   the effect of spin orbit torques on   current induced domain wall motion  has  also been addressed   in Ref.~\cite{Obata08PRB,Ryu12JMMM,Kim12PRB,Martinez12JAP,Seo12APL,Hayashi12JPCM,Martinez13AA,Martinez13APL,Stier13PRB,Martinez13IToM,Hals13PRB,Linder13PRB,Linder13PRBa,Ryu13APL,Haazen13NM}.

To conclude, we have studied the current induced DW motion in perpendicularly magnetized nanostripes in the presence of spin orbit torques. We show using micromagnetic simulations that the direction  of the current induced DW motion and the associated DW velocity depend on the relative values of the field like torque (FLT) and the Slonczewski like torques (SLT). The results are well explained by a collective coordinate model which  is used to draw a phase diagram of the DW dynamics as a function of the FLT and the SLT. We   show that  a large increase in the DW velocity can be reached by a proper tuning of both torques. This work was supported by  project Agence Nationale de la Recherche, project   ANR 11 BS10 008 ESPERADO.

\end{document}